\documentclass[a4paper,11pt]{article}
\usepackage{jinstpub} 
\usepackage{lineno}
\usepackage{multirow}
\usepackage{siunitx}
\usepackage{subfig}
\usepackage{comment}
\usepackage{lineno}

\title{\boldmath Study experimental time resolution limits of recent ASICs at Weeroc with different SiPMs and scintillators}

\author[a,1]{Tasneem Saleem,\note{Corresponding author. }}
\author[a]{Salleh Ahmad,}
\author[a]{Jean-Baptiste Cizel,}
\author[a,b]{Christophe De La Taille,}
\author[a,b]{Maxime Morenas,}
\author[c]{Vanessa Nadig,}
\author[a]{Florent Perez,}
\author[c]{Volkmar Schulz,}
\author[c,2]{Stefan Gundacker,\note{Shared last authorship.}} 
\author[a,2]{and Julien Fleury}

\affiliation[a]{Weeroc, 4 avenue de la baltique, BP 515, 91140 Villebon sur Yvette, France}
\affiliation[b]{Omega laboratory, Ecole Polytechnique CNRS-IN2P3, Palaiseau, France}
\affiliation[c]{Department of Physics of Molecular Imaging Systems, RWTH Aachen University, Pauwelsstrasse 17, 52074, Aachen, Germany}

\emailAdd{tasneem.saleem@weeroc.com}

\abstract{ Medical applications, such as Positron Emission Tomography (PET), and space applications, such as Light Detection and Ranging (LIDAR), are in need of highly specialized ASICs. Weeroc, in collaboration with different partners, is highly involved in developing a new generation of front-end ASICs. In the context of a joined LIDAR project among Weeroc, CNES, and Airbus, Weeroc is working on the development of Liroc, an ASIC for space LIDAR application. Weeroc is also working on advancing ASICs for medical applications with Radioroc under development and intended to be used for PET applications.

This study experimentally evaluates the time resolution limits of these ASICs in different configurations, with some of the most recent silicon photomultiplier (SiPM) technologies available on the market, coupled to different scintillation crystals. The best single-photon time resolution (SPTR) was achieved using FBK NUV-HD SiPMs with an FWHM of \SI{90} {\pico\second} with Liroc and \SI{73}{\pico\second} with Radioroc. Furthermore, the coincidence time resolution (CTR) of Radioroc was studied with different crystal sizes. Using a large LYSO:Ce,Ca crystal of (3$\times$3$\times$\SI{20} {\milli\metre\cubed}) with Broadcom Near UltraViolet - Metal in Trench (NUV-MT) yields a CTR of \SI{127}{\pico\second} (FWHM). The best CTR of Radioroc was determined to \SI{83}{\pico\second} (FWHM) with Broadcom NUV-MT SiPMs coupled to LYSO:Ce,Ca (2$\times$2$\times$\SI{3} {\milli\metre\cubed}) from Taiwan Applied Crystal (TAC). }


\arxivnumber{2305.14855} 

\begin{document}
\maketitle
\flushbottom

\section{Introduction}

Solid-state photo-detectors like silicon photomultipliers (SiPMs) play an important role in digitizing optical photons in medical imaging, life sciences, and high-energy physics. The use of high-performance application-specific integrated circuits (ASICs) has been a key element for getting the best performance of SiPMs and has allowed their successful use in large detectors. Novel high-performance readout electronics need to be evaluated regarding their single-photon time resolution (SPTR) achievable with state-of-the-art analog SiPMs and scintillation materials. Weeroc has developed several multi-channel chips such as Liroc and Radioroc, which allow the usage of compact photo-detectors applied in several large experiments for timing measurements. Some of the targeted applications are Positron Emission Tomography (PET) and Light Detection and Ranging (LIDAR). For both applications, the ASICs are required to have a high intrinsic timing performance in order to optimize the overall timing performance of the detection system incorporating the ASICs. The analog performance of each chip is the key element, which is required to have sufficient bandwidth (in the order of 1 GHz) accommodating the SiPM's fast signal components, low noise to reduce the signal jitter, and reasonable power consumption in order to comply with the power budget in a multi-channel detection system.

LIDAR is a laser imaging method used to create high-resolution three-dimensional representations of scanned surfaces or to determine distances to objects by measuring the two-way travel time of a laser light pulse. For instance, the space industry uses LIDAR to scan planet surfaces before landing missions, distances between spacecrafts, and for many other applications. Having the most sensitive detector is critical to measure long distances, particularly in space applications. The development of readout ASICs for SiPM-based LIDAR is relatively new and Liroc is one of the first products available on the market. It should be mentioned that other ASICs such as TOFLAR \cite{TOFLAR} for similar applications are expected to be available on the market later. 

In time-of-flight (TOF) PET applications, a gain in the signal-to-noise ratio (SNR) is directly dependent on the timing performance of the PET system, i.e., the coincidence time resolution (CTR) \cite{Surti98}.
Current state-of-the-art TOF-PET achieves a CTR of 228 ps \cite{Prenosil476}. The electronic front-end is one of the three key contributors to the overall timing performance \cite{Gundacker_2020}, which is why special care needs to be taken with respect to the electronic noise and bandwidth of the front-end. Emerging front-end designs employing high-frequency (HF) readout electronics show record-breaking timing performance \cite{Gundacker_2019, Cates_2018}, but are far from being transferred to system level due to a high power consumption and large form factor \cite{HFStefan, Cates_2022}. Hence, highly integrated front-end designs in the form of ASICs are required. It has been shown that current system-applicable solutions available on the market show a non-negligible contribution of the front-end on the measured CTR \cite{Nadig2022}. 
For PET, several ASIC solutions have been developed over the past two decades, amongst others: NINO \cite{NINO_FlexTOT2}, TOFPET2\cite{TOFPET2}, STIC\cite{STIC}, and FlexTOT\cite{NINO_FlexTOT}. Apart from NINO, all the cited ASICs are mixed-signal ASICs with embedded digital processing. Weeroc's new ASICs lineup will have both, analog and mixed-signal ASICs, in order to provide flexibility to the users.

This paper presents the latest results obtained with Liroc and Radioroc in SPTR and CTR measurements. As the time resolution is also dependent on the photosensor technology and not only on the electronics, several measurements have been conducted with several detectors (different SiPM technologies and scintillators) to provide a comprehensive overview. The best achieved SPTR and CTR values are obtained with FBK SiPMs of the  Near UltraViolet - High-Density type with a metal mask (NUV-HD-LF M3) and with the Broadcom Near UltraViolet - Metal in Trench (NUV-MT) SiPMs coupled to TAC LYSO:Ce,Ca 2$\times$2$\times$\SI{3} {\milli\metre\cubed} crystal. The timing performance of different measured CTR and SPTR of the various scintillators/SiPM/ASIC configurations are summarized and discussed.

\section{ASICs development at Weeroc: Liroc and Radioroc}

Currently, at Weeroc, different ASICs are available for reading out SiPMs, among them: Liroc and Radioroc. Other ASICs are also in development, which will incorporate signal conversion and data readout on-chip. All photo-detector readout ASICs are done using the TSMC \SI{130}{\nano\meter} process.

First LIDAR-dedicated SiPMs are currently getting on the market among other detectors such as photo-multiplier tubes (PMTs). However, other available multi-channel readout electronics do not allow for an SPTR precise enough for applications of interest. The main readout requirements of LIDAR are an excellent timing resolution and a few ns double-peak separation. None of the ASICs on the market allows for such a fast response. Weeroc has designed a LIDAR dedicated multi-channel readout chip prototype (Liroc), focusing on a high bandwidth and fast baseline return to allow fast counting and time measurements.

In scientific instrumentation, medical imaging, and radiation detection in general, SiPMs have been the number one choice due to their robustness and small form factor. Weeroc has been supplying ASICs in applications such as cosmic ray observation telescopes, Compton cameras, and PET scanners; such as the trimodal PET/MRI/EEG TRIMAGE scanner \cite{TRIMAG} based on the Triroc ASIC \cite{Triroc}. Most of these applications require a low-noise analog amplifier with a noise level better than a single photo-electron and excellent intrinsic timing jitter in the order of a few ps. Additionally, most cited applications are in need of charge measurements to reproduce the input signal spectrum and an energy cut-off feature. The Radioroc ASIC has been developed to fulfill these requirements and also will be replacing Weeroc's previous generation ASICs such as Petiroc2A \cite{Petitroc}, Triroc1A \cite{Triroc2} and Citiroc1A \cite{Citiroc}. This chip also incorporates a versatile output system so that users would be able to select the outputs based on their applications' needs, such as photon counting and energy measurement with charge integration.

\section{ASICs characterization: Materials and methods}

To evaluate the timing performance in terms of CTR and SPTR of the two ASICs Liroc and Radioroc, a set of measurements has been carried out at RWTH Aachen University. 

\subsection{Devices and characteristics}
Table \ref{tab:SiPMList} summarizes the different SiPM technologies used to carry out these measurements. The NUV-UHD SiPM type is a new SiPM technology developed at FBK that implements a new microcell border structure. The NUV-HD-LF M3 is another structure from FBK with a metal mask to shield the edges of the subcells \cite{FBK_MetalMask}. The Broadcom NUV-MT is the newest development from Broadcom and it has a special metal trench structure in the inter-pixel region \cite{BRCM_MT}.

\begin{table}[htbp]
\centering
\caption{List of SiPM Technologies used in this study}
\label{tab:SiPMList}
\smallskip
\resizebox{\columnwidth}{!}{
\begin{tabular}{|l|c|c|c|}
\hline
SiPM Producer \& Type &	SiPM Size [mm$^2$]	& Breakdown voltage [V] & Technology Specification\\
\hline
FBK NUV-HD-RH UHD-DA & 1$\times$1 &	32.5 & UHD: Ultra High Density\\
FBK NUV-HD-RH UHD-DE &	1$\times$1 & 32.5 &UHD: Ultra High Density\\	
FBK NUV-HD-LF M3 & 	1$\times$1 &	32.5 & M: Metal Mask, LF: Low Field\\
BRCM NUV-MT &	4$\times$4 &	32 & MT: Metal in Trench\\
BRCM AFBR-S4N33C013 &	3$\times$3& 27  & NUV-HD \\
HPK S13361-2050-08 &	2$\times$2 & 53 & 8$\times$8 array\\	
\hline
\end{tabular}}
\end{table}

Moreover, table \ref{tab:CrystalList} lists all scintillators that have been used in this study, i.e. scintillating (LYSO) and non-scintillating (PbF$_{2}$) crystals. The effect of codoping with Calcium (LYSO:Ce,Ca) has been studied in previous publications \cite{LYSODoping}. Co-doped scintillators are particularly interesting because of their faster scintillation rise times and decay times leading to improved timing \cite{GUNDACKERLYSO, NadigLYSO}. Additionally, non-scintillating crystals, e.g. PbF$_{2}$, with sole prompt photon emission due to the Cherenkov effect, are interesting to test the single photon response of the SiPMs and readout electronics. The Cherenkov effect has already been studied in several publications and its potential influence on the CTR was discussed \cite{Cherenkove1, Cherenkove2, Cherenkove3}.

\begin{table}[htbp]
 \centering
\caption{List of scintillator crystals coupled to SiPM used in this study.}
\label{tab:CrystalList}
\smallskip
\begin{tabular}{|l|l|c|c|c|}
\hline
Measurement Types &	Crystal Type &	Crystal Size [mm$^3$] \\
\hline
SPTR measurement &	EPIC PbF$_{2}$ (black-painted) &	2$\times$2$\times$3\\
\hline
\multirow{2}{*}{CTR Measurement} &	TAC LYSO:Ce,Ca (PMI1$\times$050/1) &	2$\times$2$\times$3\\
	& TAC LYSO:Ce,Ca (PMI1$\times$025/6) &	3$\times$3$\times$20\\
\hline
\end{tabular}
\end{table}

\subsection{Experimental setup}
\begin{figure}[htbp]
\centering
\includegraphics[width=0.7\textwidth]{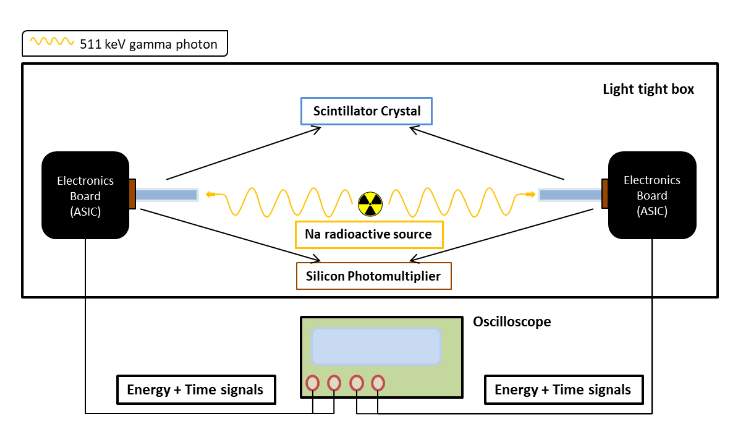}
\caption{Schematic of the experimental setup used for CTR and SPTR measurements. \label{fig:ExperimentalSetup}}
\end{figure}

A Reference and Device Under Test (DUT) SiPMs are coupled to two scintillating crystals facing each other. The SiPMs were directly connected to each ASIC. A radioactive source of Na-22 is placed at the center between the two SiPMs which emits two back-to-back 511 keV gammas that are detected in coincidence by the crystals coupled to the SiPMs. SPTR measurement is done with a similar setup but a non-scintillating crystal (black-painted PbF$_{2}$) is used at one side. In addition, an HF readout  \cite{Gundacker_2019,HFStefan} is used as a reference to perform the SPTR measurement against the ASIC under test. The HF readout of SiPMs significantly improves the measured SPTR, allowing the evaluation of the intrinsic performance of large-area devices, e.g. SiPMs, and ASICs under test. Figure \ref{fig:ExperimentalSetup} shows the schematic of the experimental setup used for these measurements.

\begin{figure}[htbp]
\centering
\includegraphics[width=0.5\textwidth, height = 7cm]{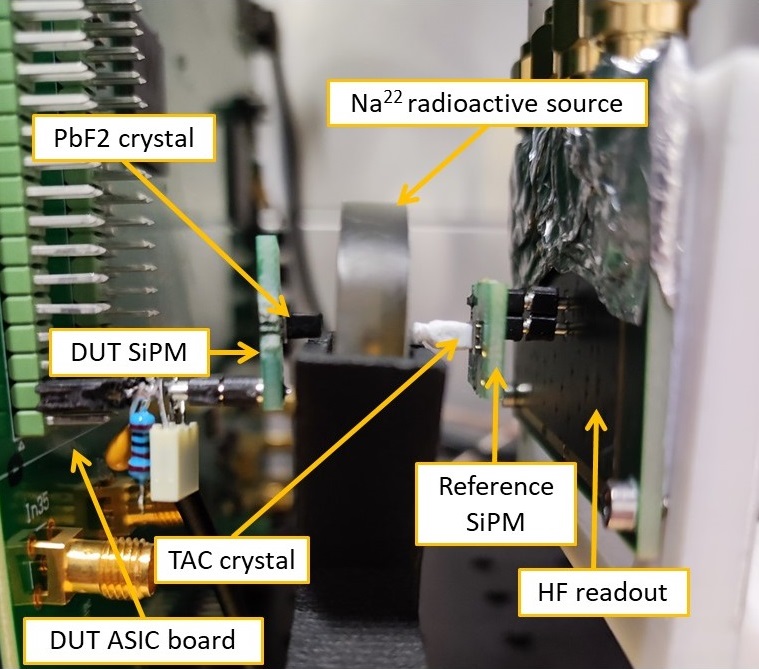}
\caption{Experimental setup used for SPTR measurements and CTR measurements. \label{fig:SetupPhoto}}
\end{figure}

The DUT SiPMs were biased with a DC power supply at several overvoltages. A positive high voltage is provided to the cathode. The positive signal collected from the anode is fed directly to the ASIC input that amplifies the SiPM signal and sends it to a discriminator. Trigger outputs of the two selected channels from the ASIC were sent to a Lecroy oscilloscope (Waverunner 9404M-MS 4 GHZ, 40 GS/s) to measure the Time over Threshold (TOT) of both the two channels and the time delay $\delta{\text{t}}$ between the two leading edges. The setup was placed in a light-tight enclosure under a fixed temperature of 16°C. Figure \ref{fig:SetupPhoto} shows a picture of the experimental setup used for SPTR and CTR measurements.

\subsection{Analysis methodology}

This section describes the method used in this study to perform analysis and determine the SPTR and CTR values. 

\subsubsection{SPTR analysis}
\begin{figure}[htbp]
\centering
\includegraphics[width=0.8\textwidth]{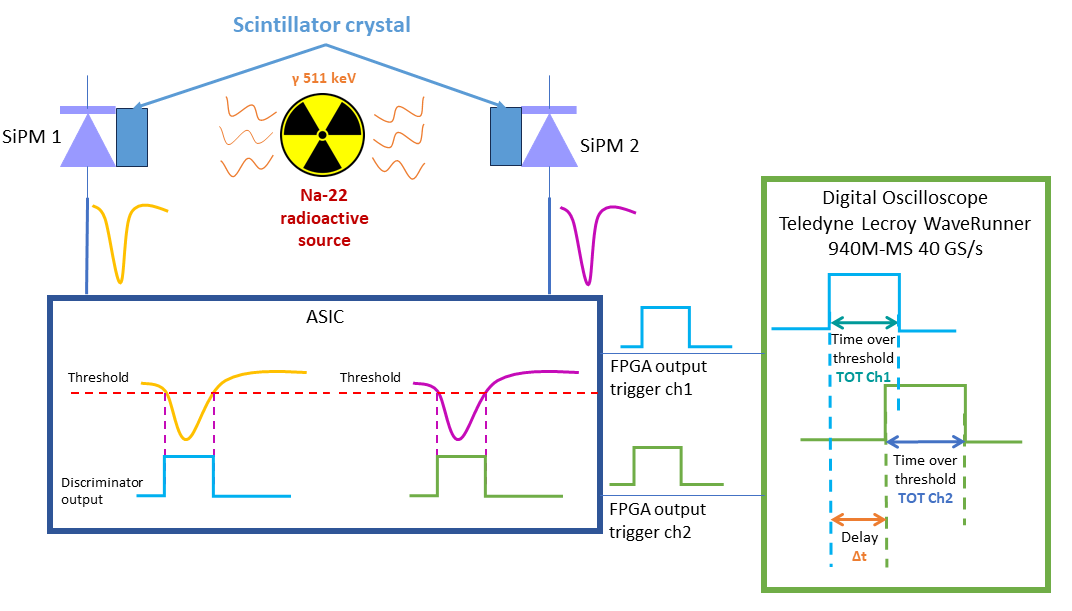}
\caption{Detailed schematic showing the different triggers and the time delay measurement. \label{fig:Trigger_DeltaT}}
\end{figure}

Time delay (i.e. the time difference) between two triggers or two leading edges time stamps is estimated in this analysis as described below. The two triggers are the trigger of the ASIC under study (Liroc or Radioroc) and the trigger of the HF reference readout. A detailed schematic of the time delay measurement and the different triggers is shown in Figure \ref{fig:Trigger_DeltaT}.

The events contained within the single photo-electron (p.e.) peak (corresponding to single-photon triggers) are selected and the histogram of the time delay signal is plotted. The full width at half maximum (FWHM) of the time delay histogram yields the SPTR.

The right end of the spectrum (i.e. delay histogram) has a tail that can be a result of delayed signals generated by photons that get converted deeper in the junction as described in \cite{FitTail1, FitTail2, FitTail3}. A Gaussian fit does not account for tails. Hence, a Gaussian convoluted with an exponential probability distribution (an exponentially modified Gaussian function) is used to fit histograms (described through equation \ref{eq2} and \ref{eq3}).

\begin{equation} \label{eq2}
{\displaystyle f(x;\mu ,\sigma ,\lambda )={\frac {\lambda }{2}}e^{{\frac {\lambda }{2}}(2\mu +\lambda \sigma ^{2}-2x)}\operatorname {erfc} \left({\frac {\mu +\lambda \sigma ^{2}-x}{{\sqrt {2}}\sigma }}\right),}
\end{equation}

where erfc is the complementary error function defined as
\begin{equation} \label{eq3}
{\begin{aligned}\operatorname {erfc}(x)&=1-\operatorname {erf}(x)\\&={\frac  {2}{{\sqrt  {\pi }}}}\int _{x}^{\infty }e^{{-t^{2}}}\,dt.\end{aligned}}
\end{equation}

All reported values of SPTR in this study are the FWHM of the fitted function. The FWHM is estimated directly from the fitted function and not from the fit parameter sigma since it bears to have no direct relationship with FWHM, unlike the Gaussian function. 

In the case of the measurement of an ASIC against the reference, the SPTR of ASICs ($\text{SPTR}_{\text{ASIC}}$) depends on the CTR value of the reference ($\text{CTR}_{\text{Ref}}$) and the time difference measured $\text{SPTR}_{\text{measured}}$. 

In this case, the SPTR is calculated as follows:

\begin{equation} \label{eq1}
 \text{SPTR}_{\text{ASIC}}=\sqrt{\text{SPTR}_{\text{measured}}^2-\frac{\text{CTR}_{\text{Ref}}^2}{2}}
\end{equation}

\noindent where $\text{CTR}_{Ref} = \SI{65}{\pico\second}$ measured using a Broadcom AFBR-S4N33C013 SiPM coupled to an LYSO:Ce,Ca 2$\times$2$\times$\SI{3} {\milli\metre\cubed} crystal.


\subsubsection{CTR analysis}

For the CTR measurement, for real coincidence, the CTR is measured using two identical boards (i.e. Radioroc against Radioroc). Hence, the two triggers are triggers of each ASIC. 

The events contained within the photoelectric absorption of the two 511 keV gammas are selected and the histogram of the time delay signal is plotted. A Gaussian is used to fit histograms and the full width at half maximum (FWHM) of the time delay histogram yields the CTR.

The CTR is corrected for the reference detector in the case of measurement against the HF reference readout. In this case, the CTR of the ASIC ($\text{CTR}_{\text{ASIC}}$) is calculated according to equation \ref{eq4}.

\begin{equation} \label{eq4}
 \text{CTR}_{\text{ASIC}}=\sqrt{2\text{CTR}_{\text{measured}}^2-\text{CTR}_{\text{Ref}}^2}
\end{equation}

\noindent The corrected CTR value takes into account the correction for the reference detector, i.e. the HF readout with a Broadcom AFBR-S4N33C013 coupled to a 2$\times$2$\times$\SI{3} {\milli\metre\cubed} LYSO:Ce,Ca crystal, with a CTR value of $\SI{65}{\pico\second}$.

\section{Measurements and results}
\subsection{Single-photon Time Resolution (SPTR) measurement}

SPTR measurements of Radioroc and Liroc have been performed against an LYSO:Ce,Ca reference detector coupled with the HF readout using several SiPMs from various manufacturers. Exposed to Cherenkov radiation by coupling the DUT SiPM to a black-painted EPIC PbF$_{2}$ 2$\times$2$\times$3 mm$^3$, the timing difference between the reference detector and the ASIC chip triggers on single-photon events is studied. 
The HF reference readout is set up with a Broadcom AFBR-S4N33C013 SiPM coupled to a TAC LYSO:Ce,Ca crystal of 2$\times$2$\times$3 mm$^{3}$. All SPTR values stated in this study have been corrected for the HF reference readout CTR. It should be mentioned that electronic noise was not subtracted for these measurements.

\begin{figure}[htbp]
\centering
\subfloat[]{\includegraphics[width=0.43\textwidth, height= 5 cm]{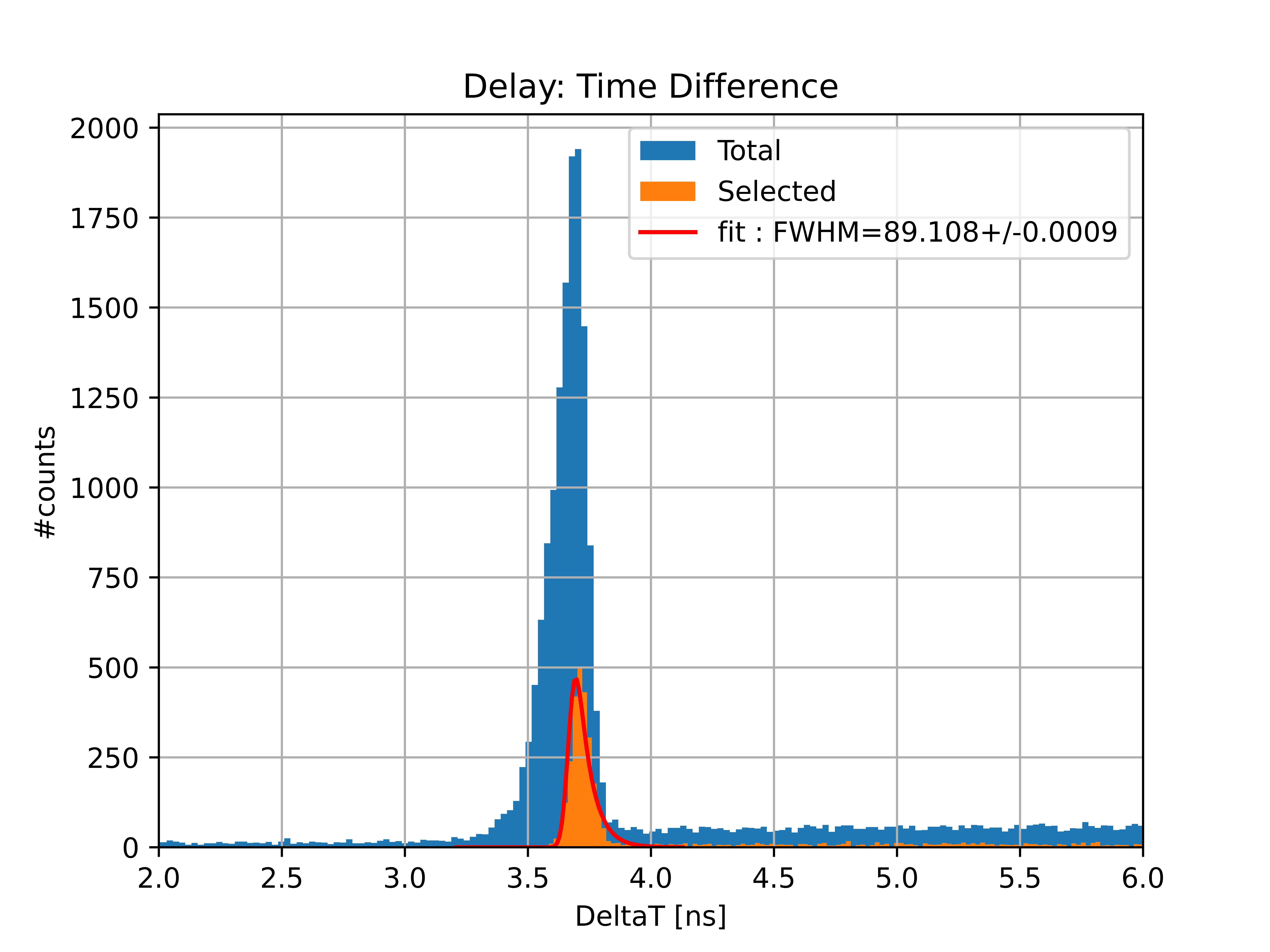}}
\qquad
\subfloat[]{\includegraphics[width=0.45\textwidth, height= 5 cm]{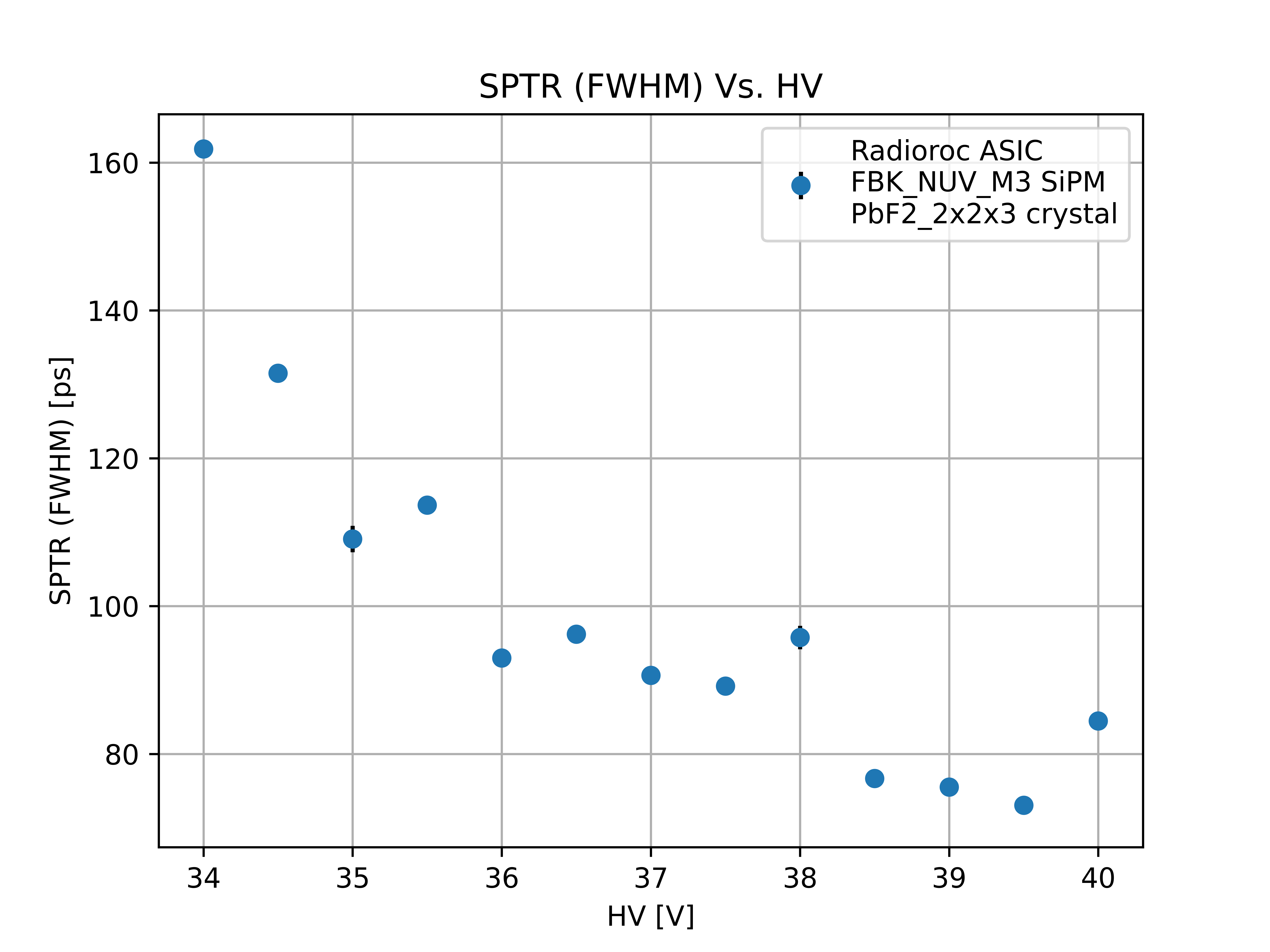}}
\caption{SPTR measurement of Radioroc with FBK NUV-HD-LF M3 SiPM coupled to Epic black-painted PbF$_{2}$ crystal of 2$\times$2$\times$3 mm$^{3}$ against the HF reference detector. (a) Histogram of the time delay between the two triggers:  all events (blue), single p.e selected events (orange), and exponentially modified Gaussian fit (red). (b) Measured SPTR of Radioroc as a function of SiPM bias voltage (HV). The best SPTR achieved is 73 ps (FWHM) at a bias voltage of 39.5 V. SPTR values are corrected to the HF reference CTR.}
\label{fig:SPTR_Radio_Results}
\end{figure}

Figure \ref{fig:SPTR_Radio_Results} shows the SPTR measurement of Radioroc with FBK NUV-HD-LF M3 SiPM against HF reference. Figure \ref{fig:SPTR_Radio_Results} (a) shows the time delay histogram, where all events are presented in blue, single p.e selected events in orange, and exponentially modified Gaussian fit in red. Figure \ref{fig:SPTR_Radio_Results} (b) shows the measured SPTR of Radioroc as a function of SiPM bias voltage (HV). The best SPTR achieved is 73 ps (FWHM) at a bias voltage of 39.5 V.

It is important to note that the values of SPTR as a function of bias voltage are plotted for a constant discriminating threshold selected by optimizing it for each bias voltage setting that gives the best timing performance for each device. Such a chosen discrimination
threshold value corresponds to about half the p.e level at that setting.

\begin{figure}[htbp]
\centering
\subfloat[]{\includegraphics[width=0.43\textwidth, height= 5 cm]{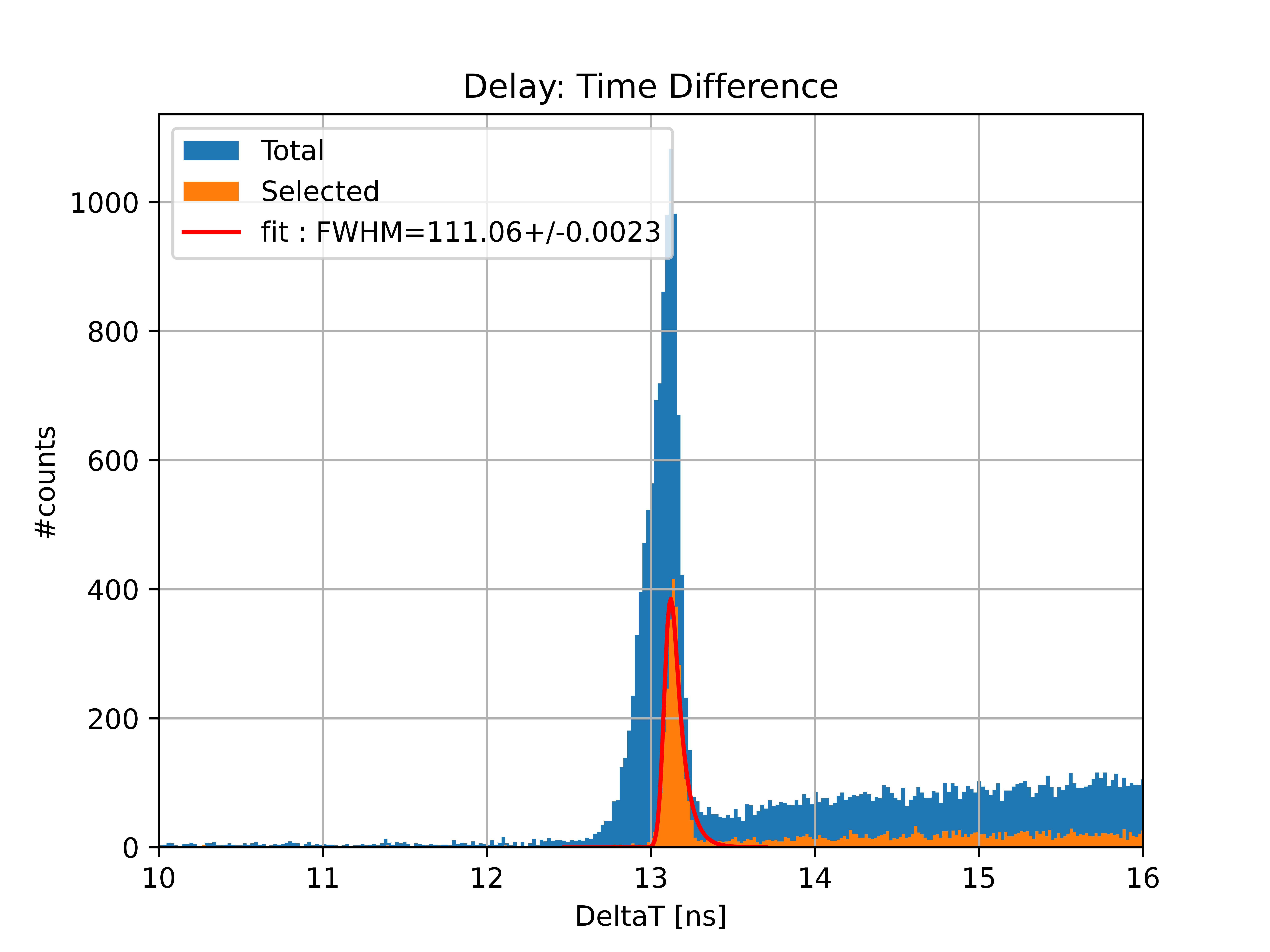}}
\qquad
\subfloat[]{\includegraphics[width=0.45\textwidth, height= 5 cm]{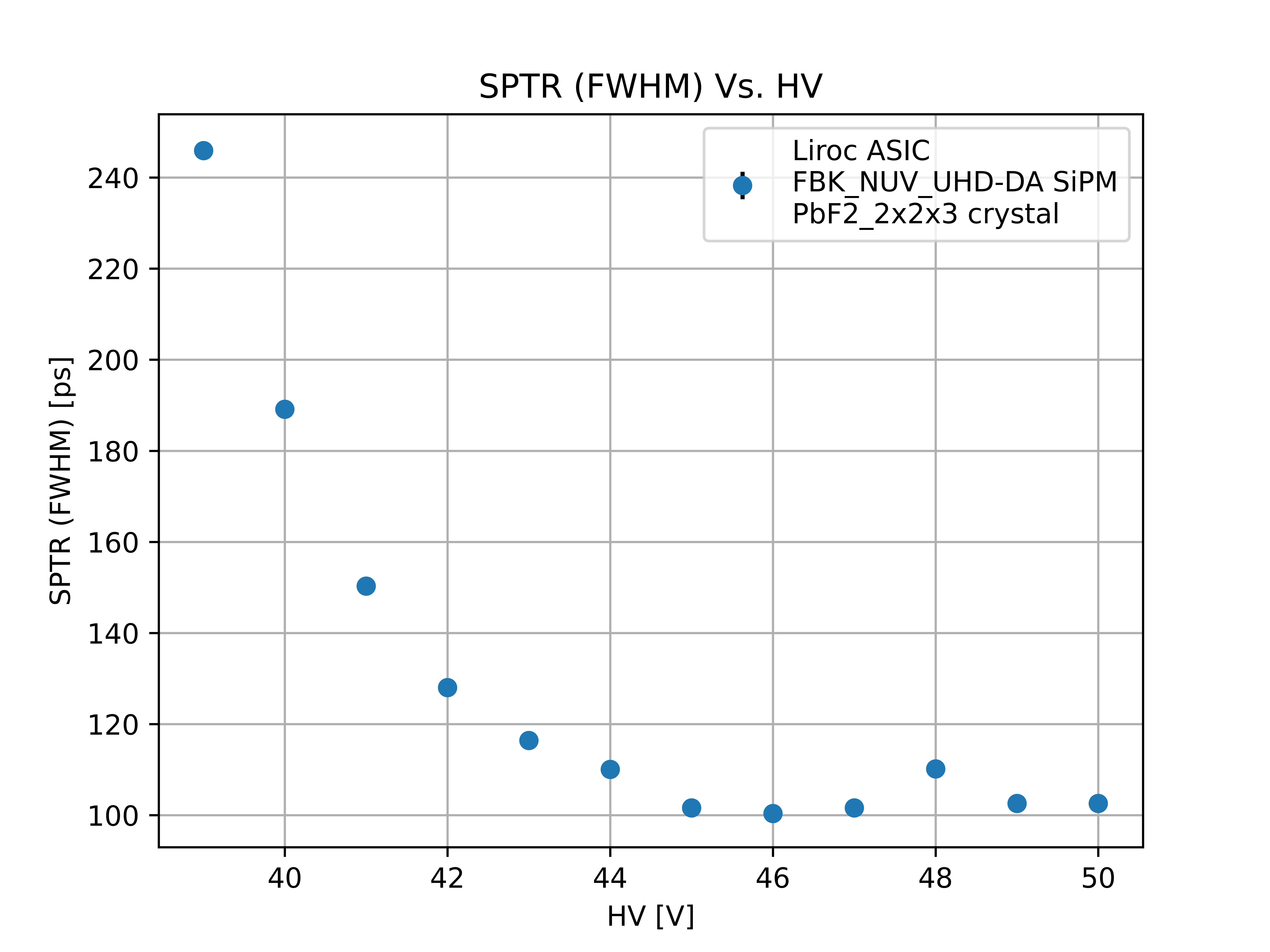}}
\caption{ SPTR measurement of Liroc with FBK NUV-HD-RH UHD-DA SiPM coupled to Epic black-painted PbF$_{2}$ crystal of 2$\times$2$\times$3 mm$^{3}$ against the HF reference detector. (a) Histogram of the time delay between the two triggers:  all events (blue), single p.e selected events (orange), and exponentially modified Gaussian fit (red). (b) Measured SPTR of Liroc as a function of SiPM bias voltage (HV). The best SPTR achieved is 102 ps (FWHM) at a bias voltage of 45 V. SPTR values are corrected to the HF reference CTR.}
\label{fig:SPTR_LIROC_FBK_DA}
\end{figure}

Figure \ref{fig:SPTR_LIROC_FBK_DA} shows the SPTR measurement of Liroc with FBK NUV-HD-RH UHD-DA SiPM against the HF reference detector. Figure \ref{fig:SPTR_LIROC_FBK_DA} (a) shows the time delay histogram, where all events are presented in blue, single p.e selected events in orange, and exponentially modified Gaussian fit in red. Figure \ref{fig:SPTR_LIROC_FBK_DA} (b) shows the measured SPTR of Liroc as a function of SiPM bias voltage (HV). The best SPTR achieved is 102 ps (FWHM) at a bias voltage of 45 V.

\begin{figure}[h]
\centering
\subfloat[]{\includegraphics[width=0.43\textwidth, height= 5 cm]{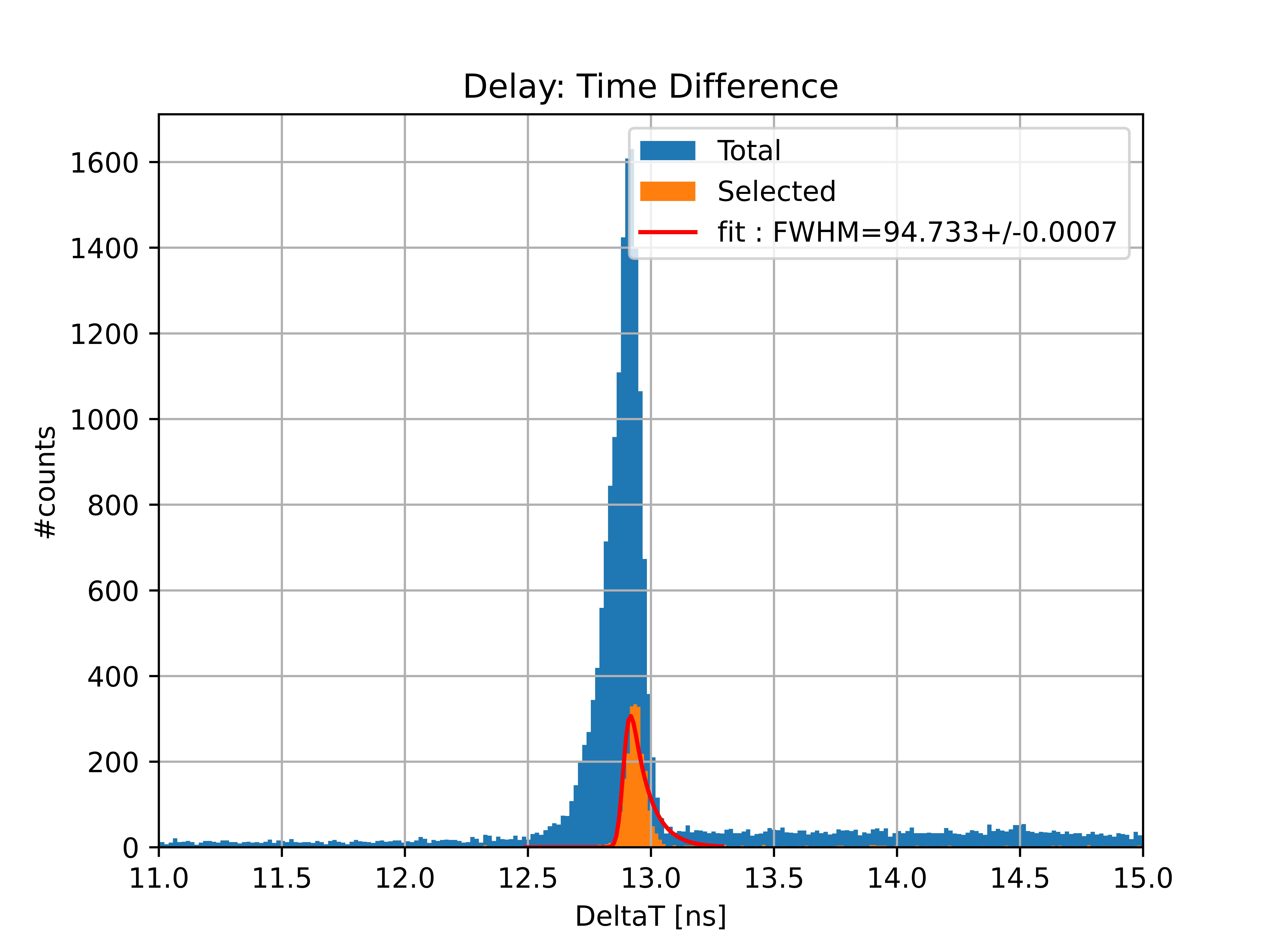}}
\qquad
\subfloat[]{\includegraphics[width=0.45\textwidth, height= 5 cm]{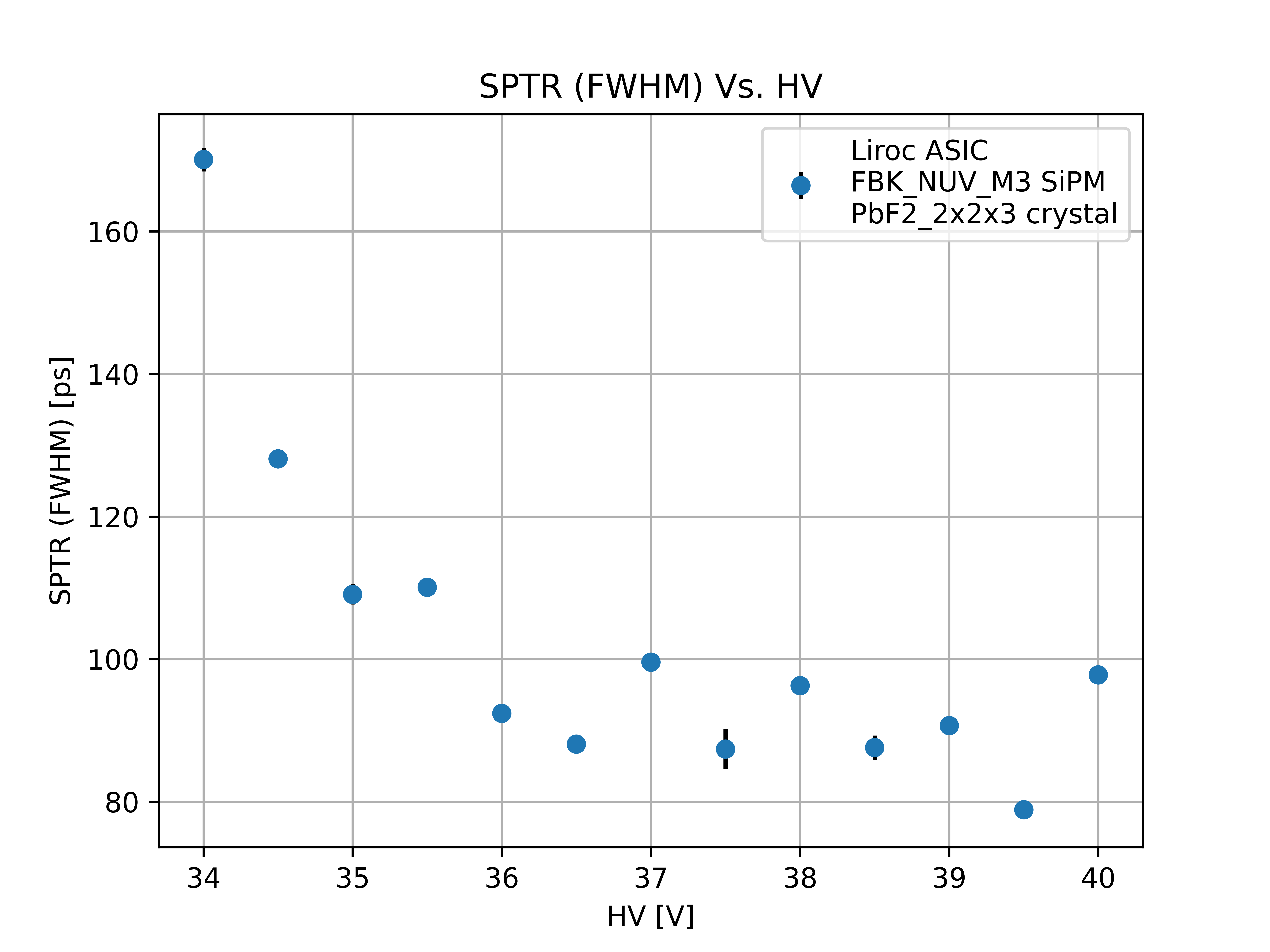}}
\caption{SPTR measurement of Liroc with FBK NUV-HD-LF M3 SiPM coupled to Epic black-painted PbF$_{2}$ crystal of 2$\times$2$\times$3 mm$^{3}$ against the HF reference detector. (a) Histogram of the time delay between the two triggers:  all events (blue), single p.e selected events (orange), and exponentially modified Gaussian fit (red). (b) The best SPTR achieved is 90 ps (FWHM) at a bias voltage of 39 V. SPTR values are corrected to the HF reference CTR.}
\label{fig:SPTR_LIROC_FBK_M3}
\end{figure}

Figure \ref{fig:SPTR_LIROC_FBK_M3} shows the SPTR measurement of Liroc with FBK NUV-HD-LF M3 SiPM against the HF reference detector. Figure \ref{fig:SPTR_LIROC_FBK_M3} (a) shows the time delay histogram, where all events are presented in blue, single p.e selected events in orange, and exponentially modified Gaussian fit in red. Figure \ref{fig:SPTR_LIROC_FBK_M3} (b) shows the measured SPTR of Liroc as a function of SiPM bias voltage (HV). The best SPTR achieved is 90 ps (FWHM) at a bias voltage of 39 V.

\begin{figure}[h]
\centering
\subfloat[]{\includegraphics[width=0.43\textwidth, height= 5 cm]{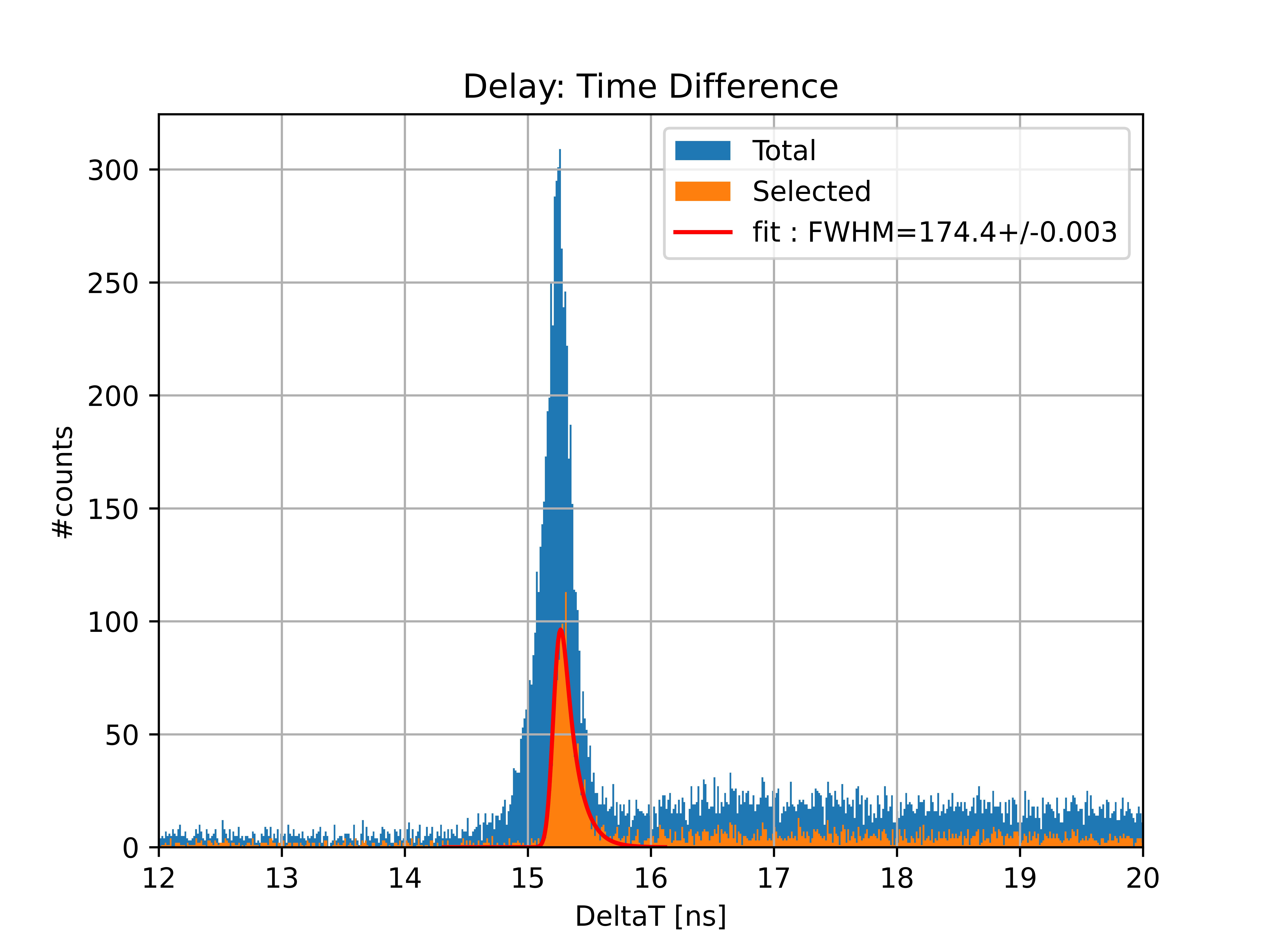}}
\qquad
\subfloat[]{\includegraphics[width=0.45\textwidth, height= 5 cm]{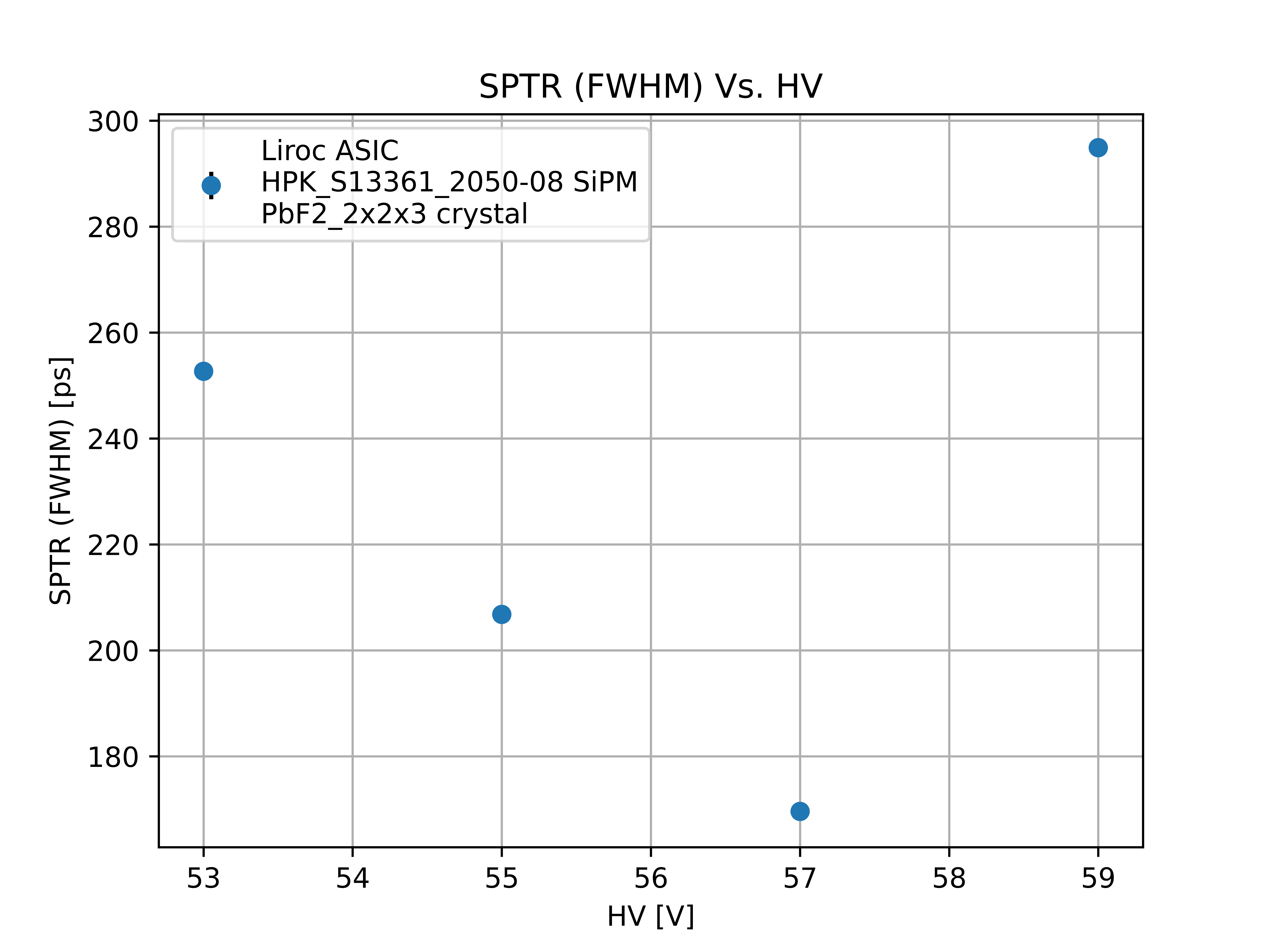}}
\caption{SPTR measurement of Liroc with Hamamatsu (S13361-2050-08) SiPM coupled to Epic black-painted PbF$_{2}$ crystal of 2$\times$2$\times$3 mm$^{3}$ against the HF reference detector. (a) Histogram of the time delay between the two triggers:  all events (blue), single p.e selected events (orange), and exponentially modified Gaussian fit (red). (b) Measured SPTR of Liroc as a function of SiPM bias voltage (HV). The best SPTR achieved is 169 ps (FWHM) at a bias voltage of 57 V. SPTR values are corrected to the HF reference CTR.}
\label{fig:SPTR_LIROC_HPK}
\end{figure}

Figure \ref{fig:SPTR_LIROC_HPK} shows the SPTR measurement of Liroc with Hamamatsu (S13361-2050-08) SiPM against the HF reference detector. Figure \ref{fig:SPTR_LIROC_HPK} (a) shows the time delay histogram, where all events are presented in blue, single p.e selected events in orange, and exponentially modified Gaussian fit in red. Figure \ref{fig:SPTR_LIROC_HPK} (b) shows the measured SPTR of Liroc as a function of SiPM bias voltage (HV). The best SPTR achieved is 169 ps (FWHM) at a bias voltage of 57 V.

\subsection{Coincidence Time Resolution (CTR) measurement}

The CTR between two SiPM-crystal coupled detector blocks, using the Radioroc ASIC is studied in this section. 

\begin{figure}[htbp]
\centering
\subfloat[]{\includegraphics[width=0.43\textwidth, height= 5 cm]{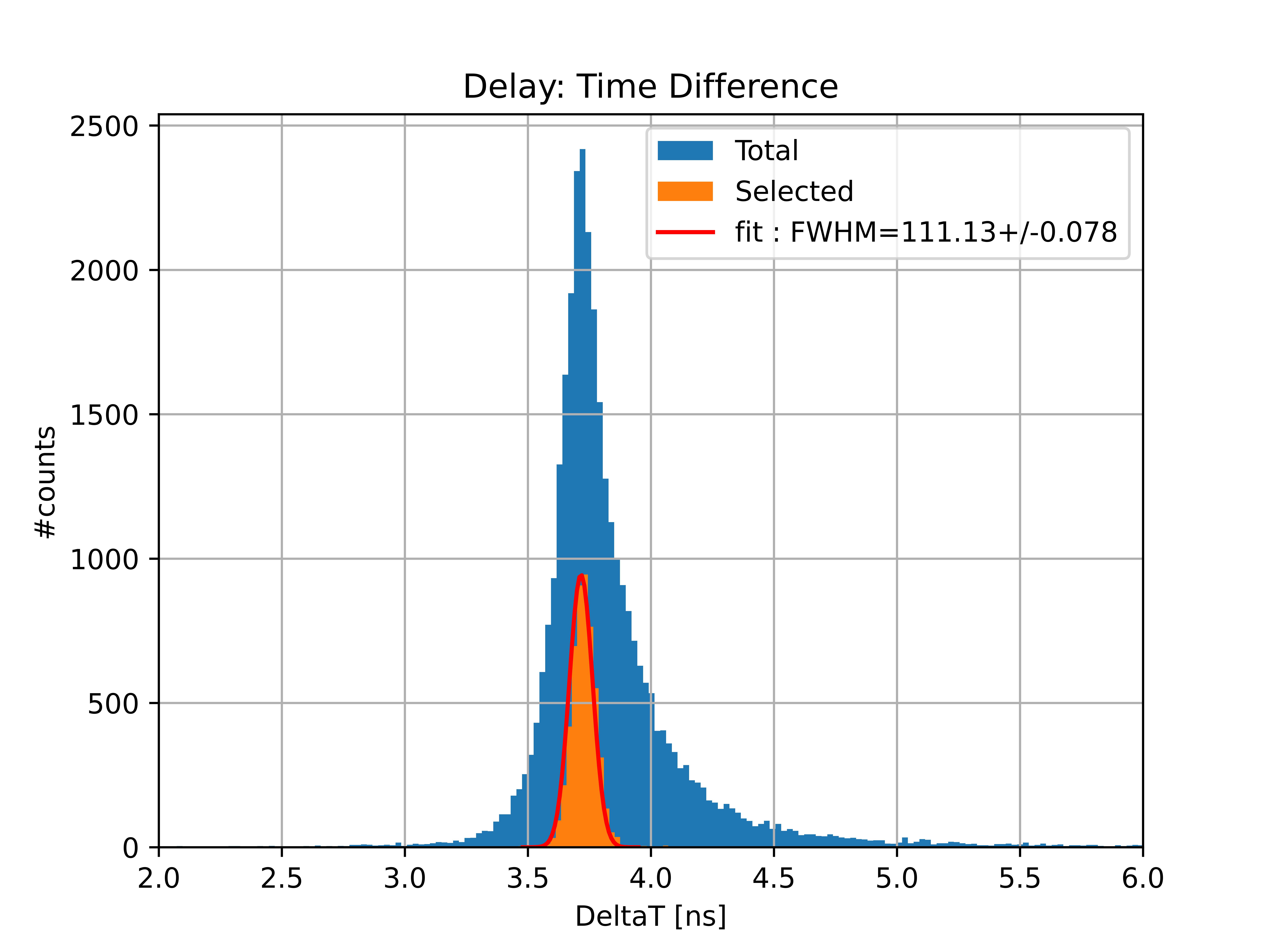}}
\qquad
\subfloat[]{\includegraphics[width=0.45\textwidth, height= 5 cm]{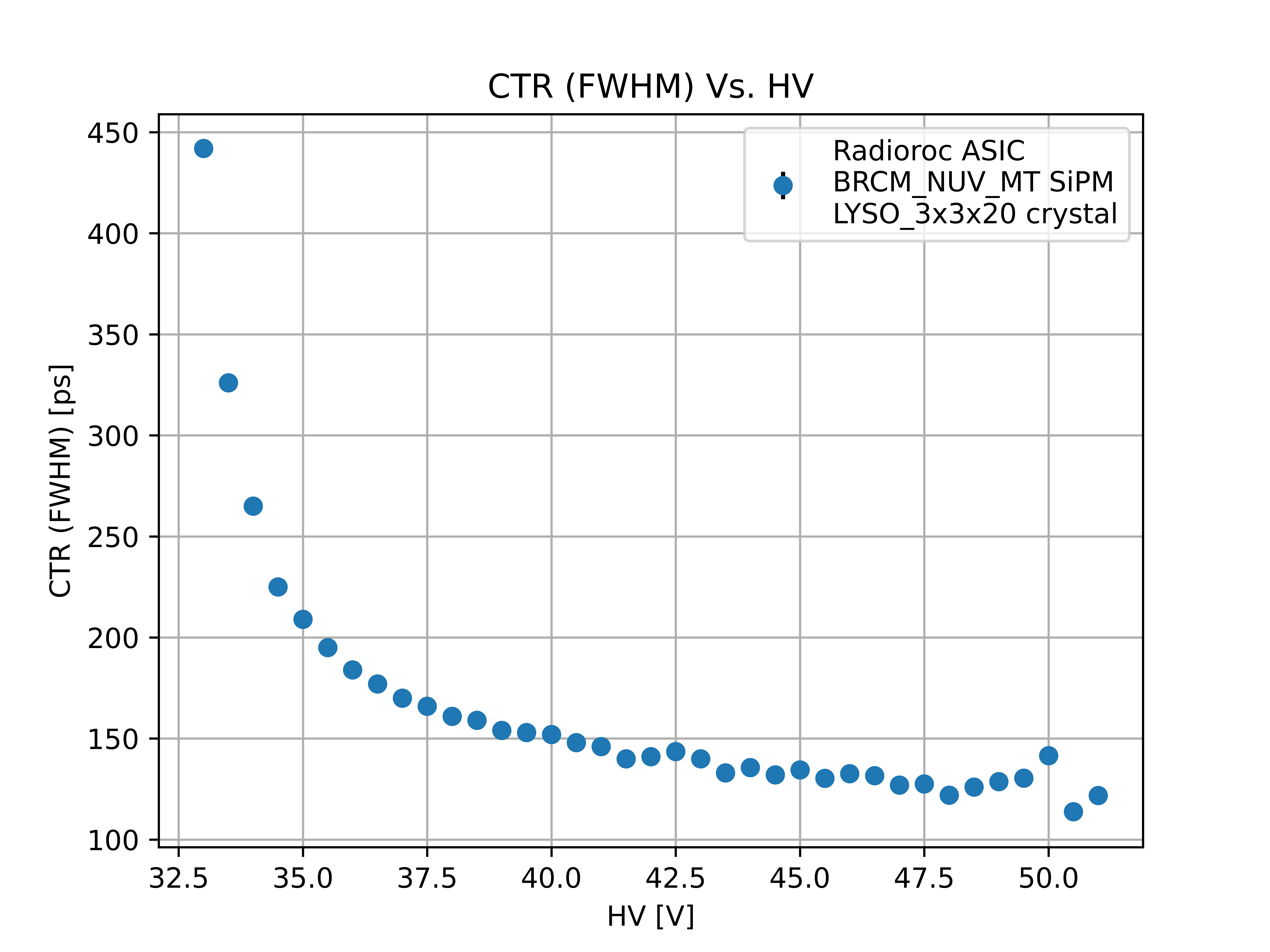}}
\caption{CTR measurement of Radioroc against Radioroc with two BRCM NUV-MT SiPM coupled to LYSO crystal of 3$\times$3$\times$20 mm$^{3}$. (a) Histogram of the time delay between the two triggers:  all events (blue), selected photopeak events (orange), and Gaussian fit (red). (b) Measured CTR of Radioroc as a function of SiPM bias voltage (HV). The best CTR of 127 ps (FWHM) at a bias voltage of 47 V.}
\label{fig:CTR_Radio_Results}
\end{figure}

Figure \ref{fig:CTR_Radio_Results} shows the CTR measurement of Radioroc against Radioroc, with two Broadcom NUV-MT SiPMs measured in coincidence. An LYSO crystal of 3$\times$3$\times$20 mm$^{3}$ was coupled to both SiPMs. Figure \ref{fig:CTR_Radio_Results} (a) shows the time delay histogram, where all events are presented in blue, selected photopeak events in orange, and Gaussian fit in red. Figure \ref{fig:CTR_Radio_Results} (b) shows the measured CTR of Radioroc as a function of SiPM bias voltage (HV). The best CTR achieved is 127 ps (FWHM) at a bias voltage of 47 V.

\begin{figure}[htbp]
\centering
\subfloat[]{\includegraphics[width=0.43\textwidth, height= 5 cm]{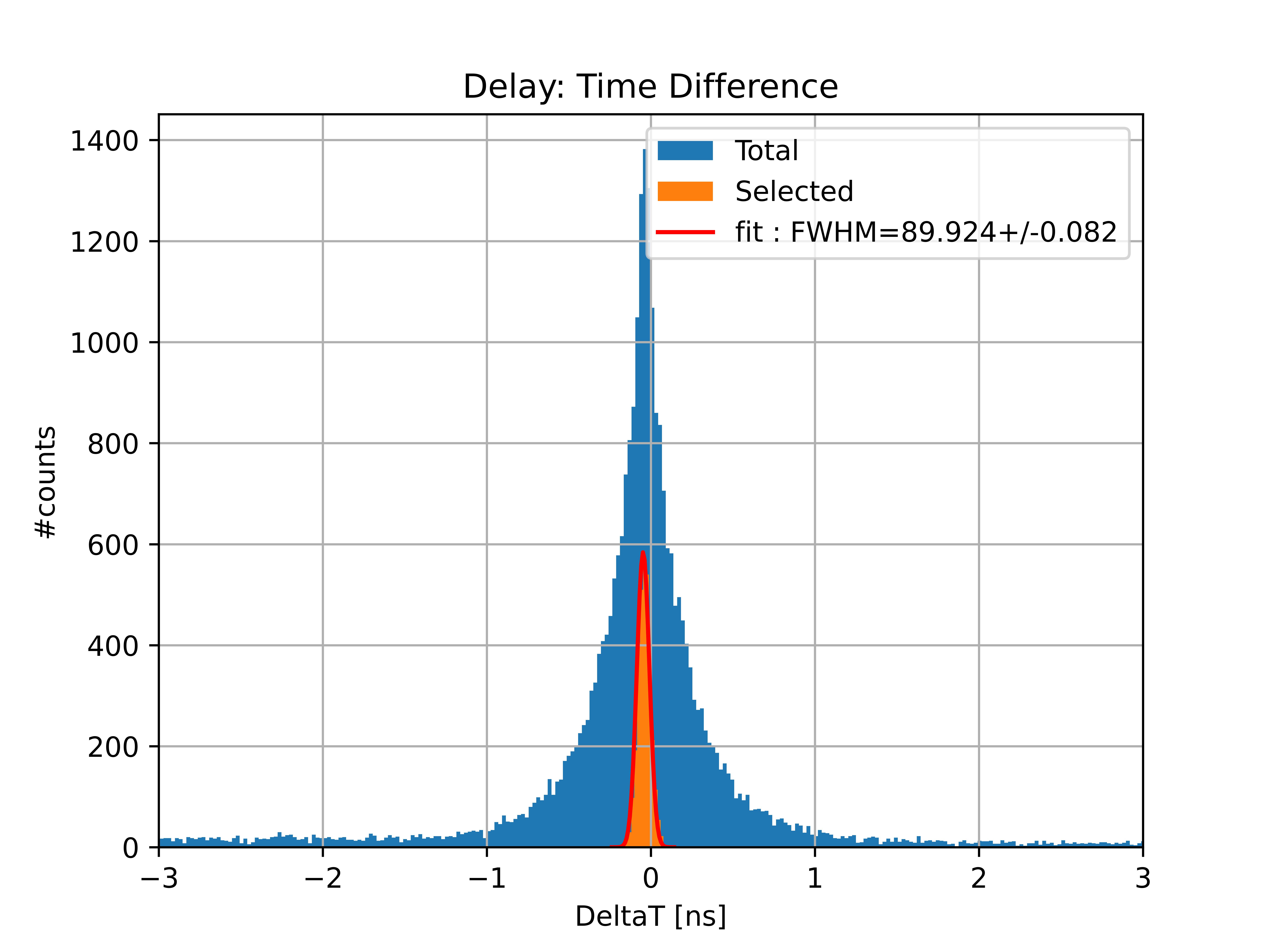}}
\qquad 
\subfloat[]{\includegraphics[width=0.45\textwidth, height = 5 cm]{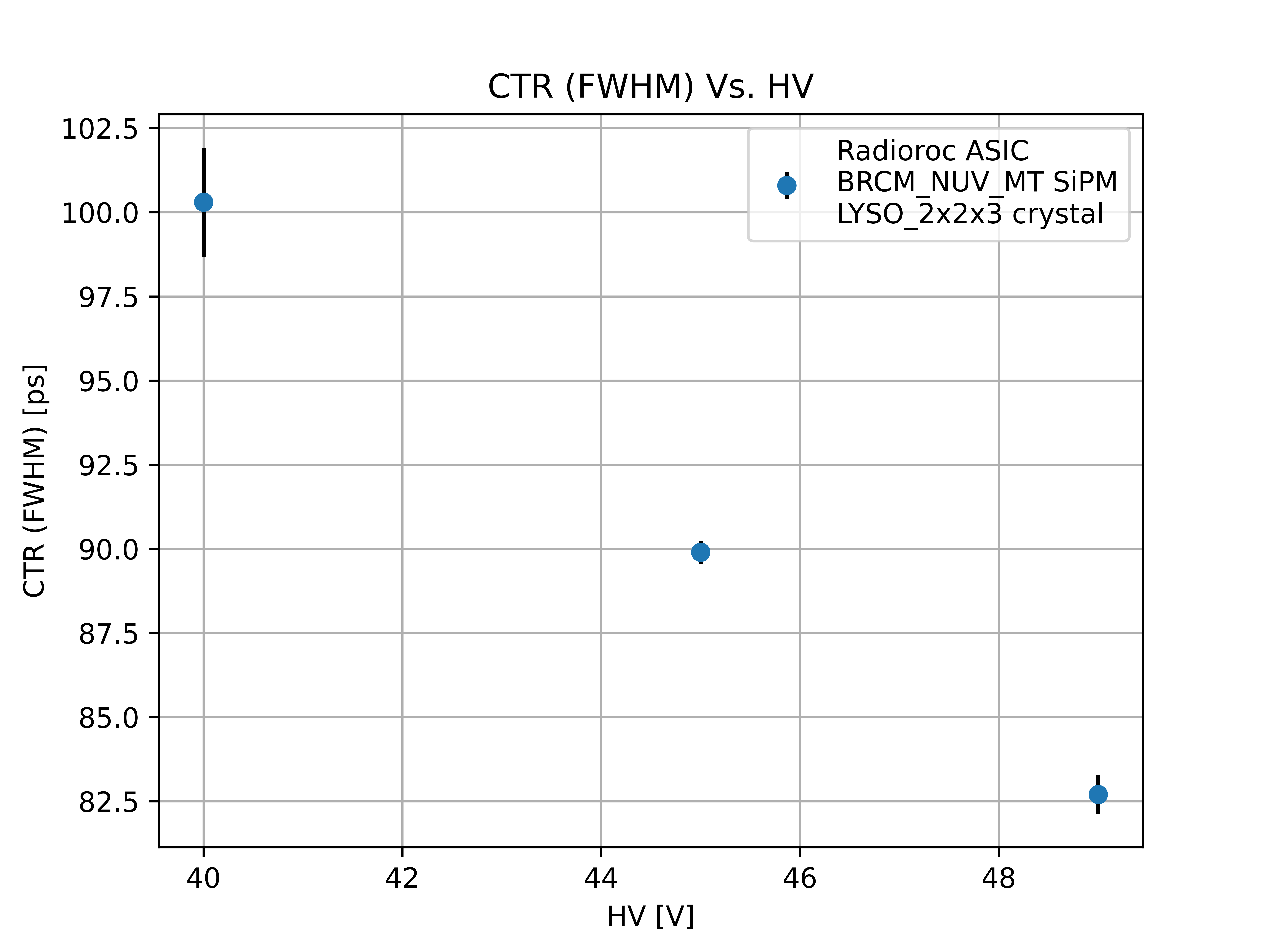}}
\caption{CTR measurement of Radioroc against Radioroc with two Broadcom NUV-MT SiPMs are measured in coincidence. Both BRCM NUV-MT SiPM coupled to LYSO crystal of 2$\times$2$\times$3 mm$^{3}$. (a) Histogram of the time delay between the two triggers:  all events (blue), selected photopeak events (orange), and Gaussian fit (red). (b) Measured CTR of Radioroc as a function of SiPM bias voltage (HV). The best CTR of 83 ps (FWHM) at a bias voltage of 49 V. }
\label{fig:CTR_Radio_smallCrystal_Results}
\end{figure}

On the other hand, using smaller crystal exhibit better CTR. Figure \ref{fig:CTR_Radio_smallCrystal_Results} shows the CTR measurement of Radioroc against Radioroc, with two Broadcom NUV-MT SiPMs measured in coincidence. An LYSO crystal of 2$\times$2$\times$3 mm$^{3}$ was coupled to the SiPM. Figure \ref{fig:CTR_Radio_smallCrystal_Results} (a) shows the time delay histogram, where all events are presented in blue, selected photopeak events in orange, and Gaussian fit in red. Figure \ref{fig:CTR_Radio_smallCrystal_Results} (b) shows the measured CTR of Radioroc as a function of SiPM bias voltage (HV). The best CTR achieved is 83 ps (FWHM) at a bias voltage of 49 V.

\subsection{Energy measurement with Radioroc in ToT mode}

Fast timing and energy-resolving capabilities are crucial for a SiPM detector readout. Time-over-threshold (ToT) offers an attractive method for combining timing and energy encoding in one signal.  In the ToT method, one measures the signal pulse width at a selected threshold, which can be used to estimate the signal’s charge. And as mentioned before, Radiroc is a multi-purpose readout chip for SiPM detectors. It can measure time and energy simultaneously by employing time-of-arrival (ToA) and time-over-threshold (ToT) techniques.

\begin{figure}[htbp]
\centering
\subfloat[]{\includegraphics[width=0.43\textwidth, height= 5.5 cm]{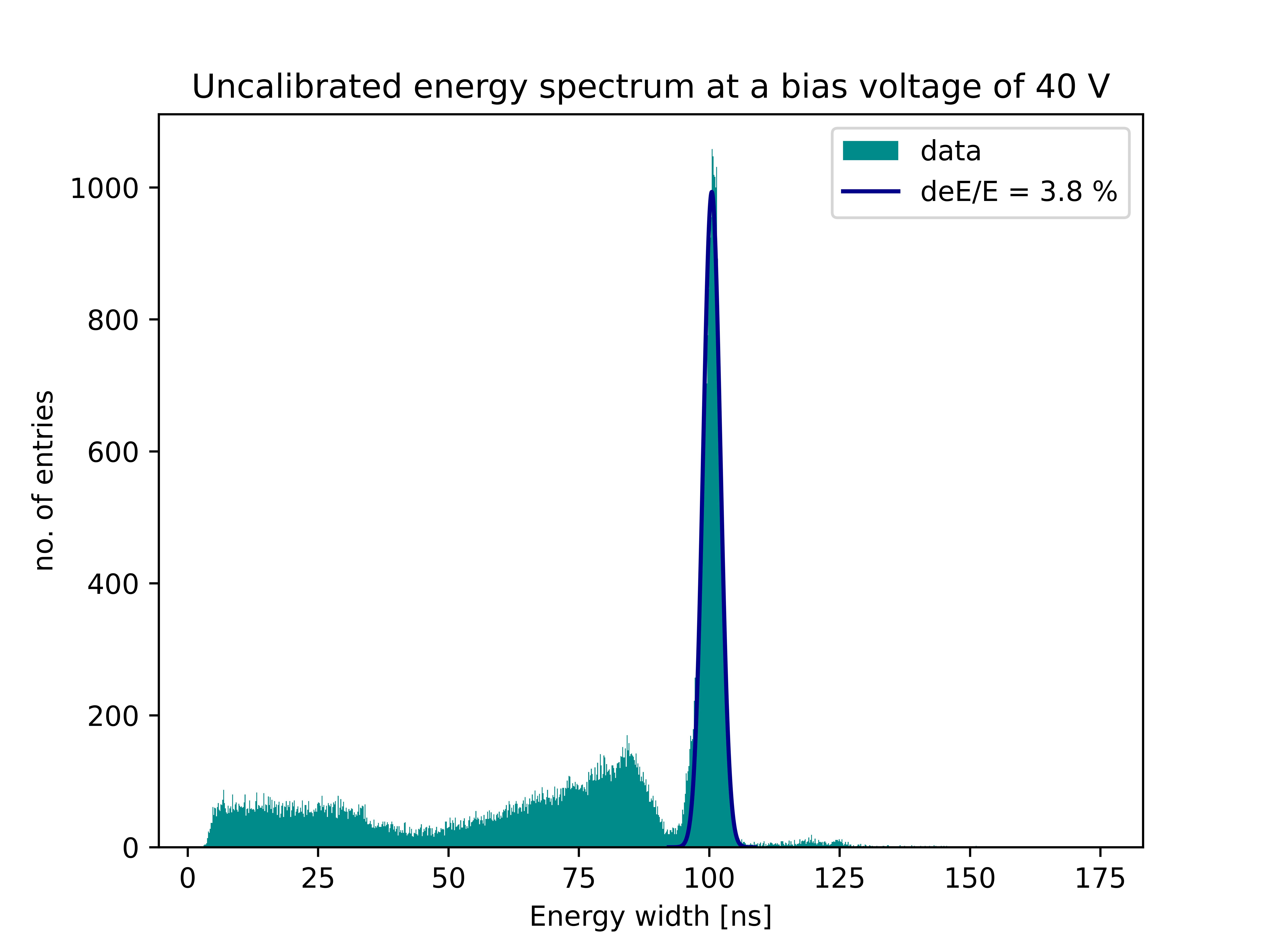}}
\qquad
\subfloat[]{\includegraphics[width=0.45\textwidth, height= 5.2 cm]{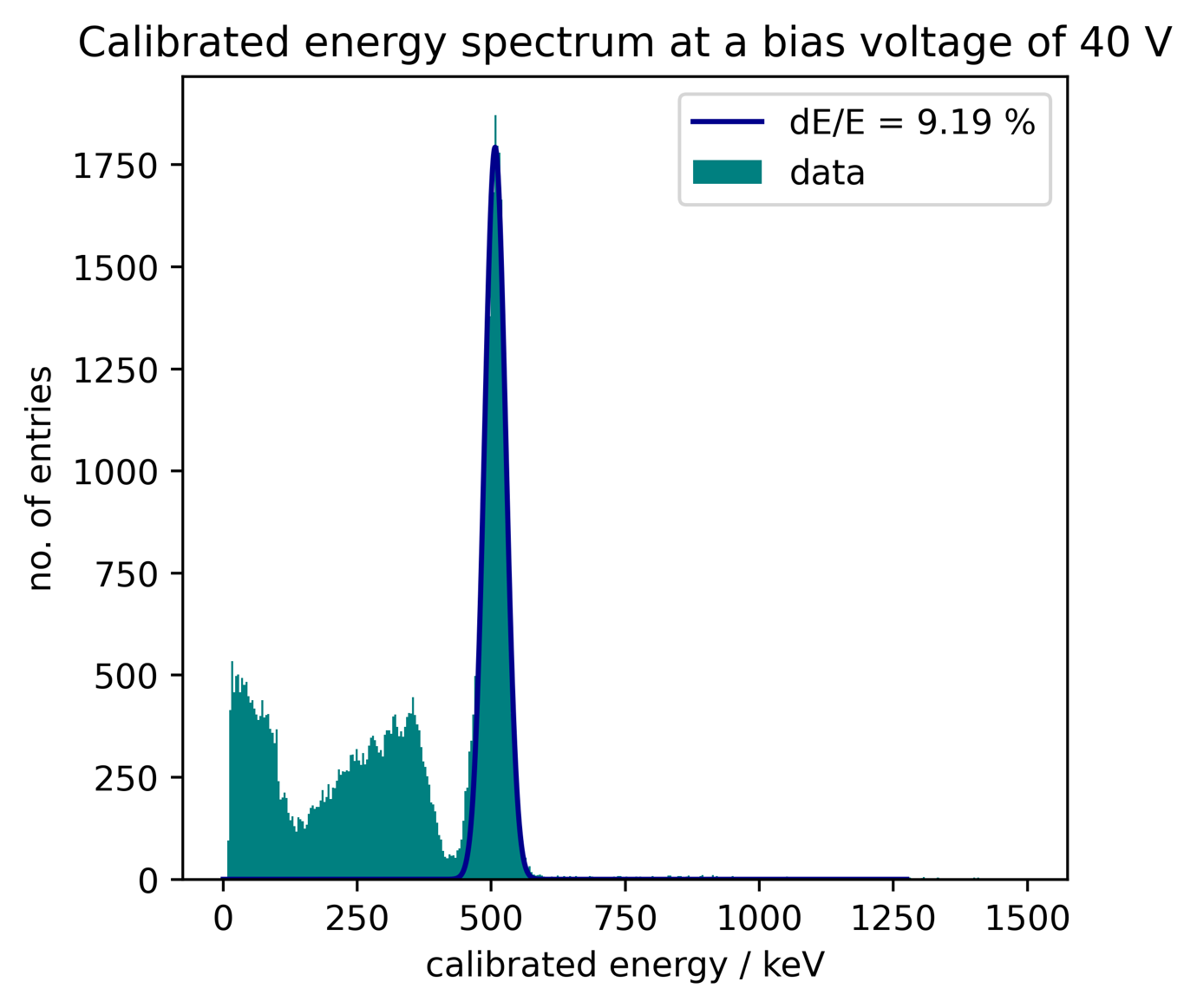}}
\caption{An example of a ToT measurement performed with Radioroc with Broadcom NUV-MT SiPM coupled to LYSO:Ce,Ca TAC crystal of 3$\times$3$\times$19 mm$^3$ at 40 V. (a) Energy spectrum before calibration. (b) Energy spectrum after calibration with an energy resolution of 9.2 \%. }
\label{fig:EnergySpectrum_TOT}
\end{figure}

In this section, an example of the energy measurement performed using Radioroc is presented and energy resolution is estimated. Figure \ref{fig:EnergySpectrum_TOT} (a) shows the energy spectrum before calibration. An energy calibration, that takes into account the saturation effects of the SiPM, was performed based on the method explained in \cite{EnergyCalib}. This method makes use of the two peaks at 511 keV and 1275 keV present in the Na-22 spectrum. The two peaks are fitted with a Gaussian to determine their centroid in time units (ns). Assigning these positions to their respective value in keV and assuming that 0 keV is located at zero, a logarithmic function is fit to these data points. The resulting parameters are used to correct the spectrum for the saturation of the SiPM and convert the time units (ns) in keV. Using a Gaussian fit, the energy resolution can then be extracted as the FWHM of the photopeak in the saturation-corrected spectrum. Figure \ref{fig:EnergySpectrum_TOT} (b) shows the saturation-corrected energy spectrum. This energy spectrum corresponds to a ToT measurement performed with Radioroc and a Broadcom SiPM (NUV-MT) coupled to a TAC LYSO:Ce,Ca crystal of 3$\times$3$\times$20 mm$^3$ at 40 V (equivalent to 8 V overvoltage). After calibration, the energy resolution at 511 keV is 9.2\%.

\section{Discussion and summary}

The SPTR and CTR of Radioroc and Liroc have been evaluated by exploring different SiPM and crystals in different configurations. Tables \ref{tab:MeasurementSummary_1} and \ref{tab:MeasurementSummary_2} below summarize all measurements carried out during this campaign providing the best time resolution value obtained. Radioroc and Liroc show an improved performance compared to other state-of-the-art circuits \cite{NINO_FlexTOT}.

\begin{table}[htbp]
 \caption{Summary of best obtained SPTR for Liroc and Radioroc in different configurations}
\label{tab:MeasurementSummary_1}
 \centering
 \resizebox{\columnwidth}{!}{
 \smallskip
 \begin{tabular}{|l|l|l|l|l|l|l|c|}
\hline
	\# & ASIC (1)	& Scintillator Crystal &	ASIC (2) &	Scintillator Crystal & 	SiPM Array & Specifications	& SPTR (FWHM) [ps]\\
\hline
1 & Radioroc & PbF$_{2}$-Black - 2$\times$2$\times$3 & HF & TAC2$\times$2$\times$3 - LYSO:Ce,ca & FBK NUV-HD-LF M3 & 39.5 V & 73\\

2 & Liroc & PbF$_{2}$-Black - 2$\times$2$\times$3 & HF & TAC2$\times$2$\times$3 - LYSO:Ce,ca & FBK NUV-HD-RH UHD-DA & 45 V & 102\\

3 & Liroc & PbF$_{2}$-Black - 2$\times$2$\times$3 & HF & TAC2$\times$2$\times$3 - LYSO:Ce,ca & FBK NUV-HD-LF M3 & 39 V & 90\\

4 & Liroc & PbF$_{2}$-Black - 2$\times$2$\times$3 & HF & TAC2$\times$2$\times$3 - LYSO:Ce,ca & HPKS13361-2050-08 & 57 V & 169\\

\hline 
\end{tabular}
}
\end{table}

\begin{table}[htbp]
\caption{Summary of best obtained CTR for Liroc and Radioroc in different configurations}
\label{tab:MeasurementSummary_2}
\centering
\resizebox{\columnwidth}{!}{
\smallskip
\begin{tabular}{|l|l|l|l|l|l|l|c|}
\hline
	\#& ASIC (1)	& Scintillator Crystal &	ASIC (2) &	Scintillator Crystal & 	SiPM Array & Specifications	& CTR (FWHM) [ps]\\
\hline
1 & Radioroc & PMI1X026-LYSO:Ce,ca – 3$\times$3$\times$20	& HF & TAC2$\times$2$\times$3 - LYSO:Ce,ca & BRCM-NUV-MT & 48 V & 137\\

2 & Radioroc &	PMI1X050-LYSO:Ce,ca – 2$\times$2$\times$3 & Radioroc	& PMI1X051-LYSO:Ce,ca - 2$\times$2$\times$3 &	BRCM-NUV-MT & 49 V & 83\\

3 & Radioroc & 	PMI1X025-LYSO:Ce,ca – 3$\times$3$\times$20	& Radioroc &	PMI1X026-LYSO:Ce,ca – 3$\times$3$\times$20& BRCM-NUV-MT & 47 V& 127\\

4 & Radioroc &	PMI1X025-LYSO:Ce,ca – 3$\times$3$\times$20 &	Radioroc & PMI1X026-LYSO:Ce,ca – 3$\times$3$\times$20 & BRCM-AFBR-S4N33C013 & 38 V & 148\\

\hline
\end{tabular}
}
\end{table}

\begin{figure}[htbp]
\centering
\includegraphics[width=.7\textwidth]{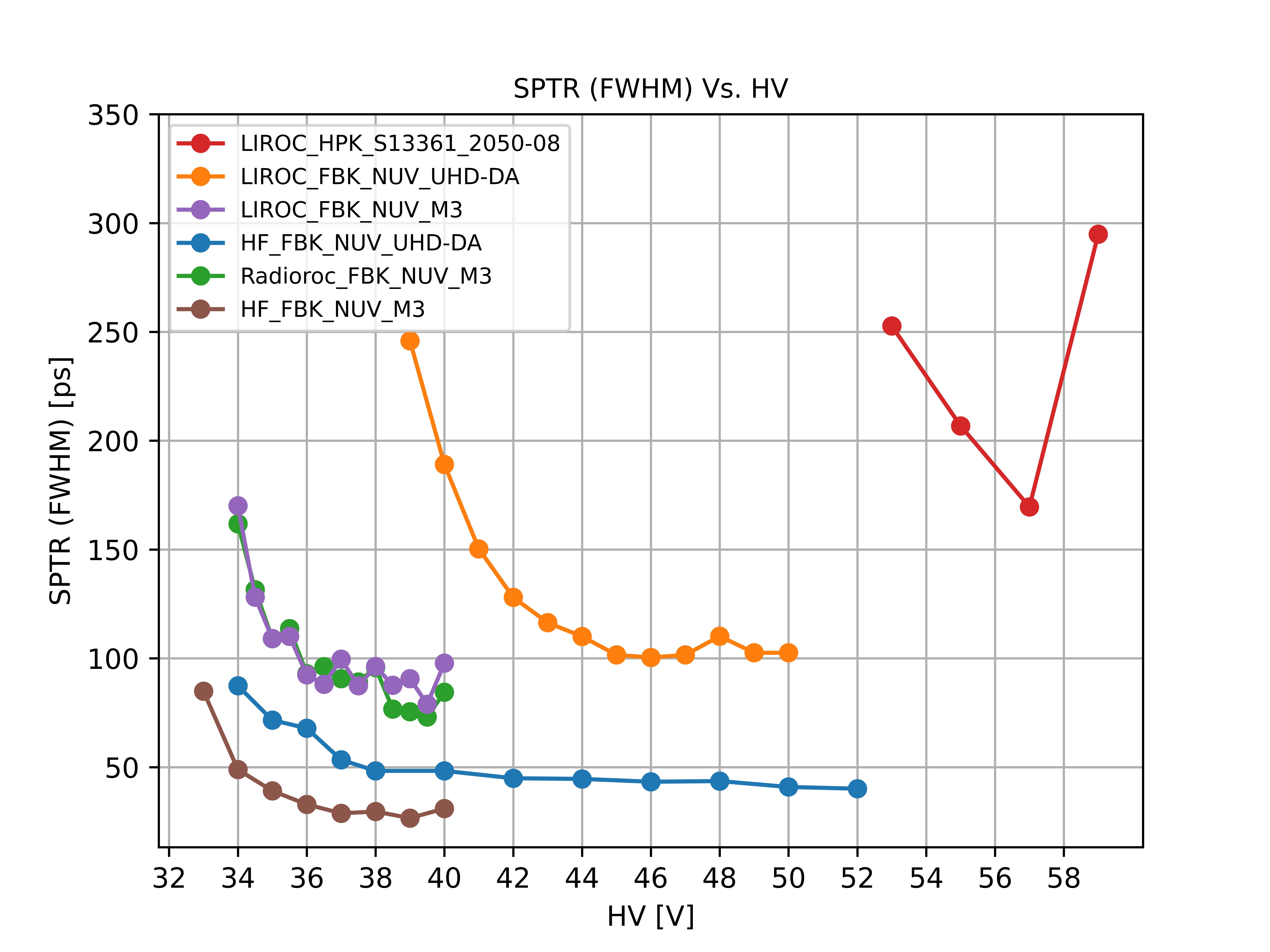}
\caption{Measured SPTR of Radioroc, Liroc, and HF reference readout with different SiPMs coupled to black-painted PbF$_{2}$ crystal of 2$\times$2$\times$3 mm$^{3}$.\label{fig:SPTR_SummaryPlot}}
\end{figure}

\begin{figure}[htbp]
\centering
\includegraphics[width=.7\textwidth]{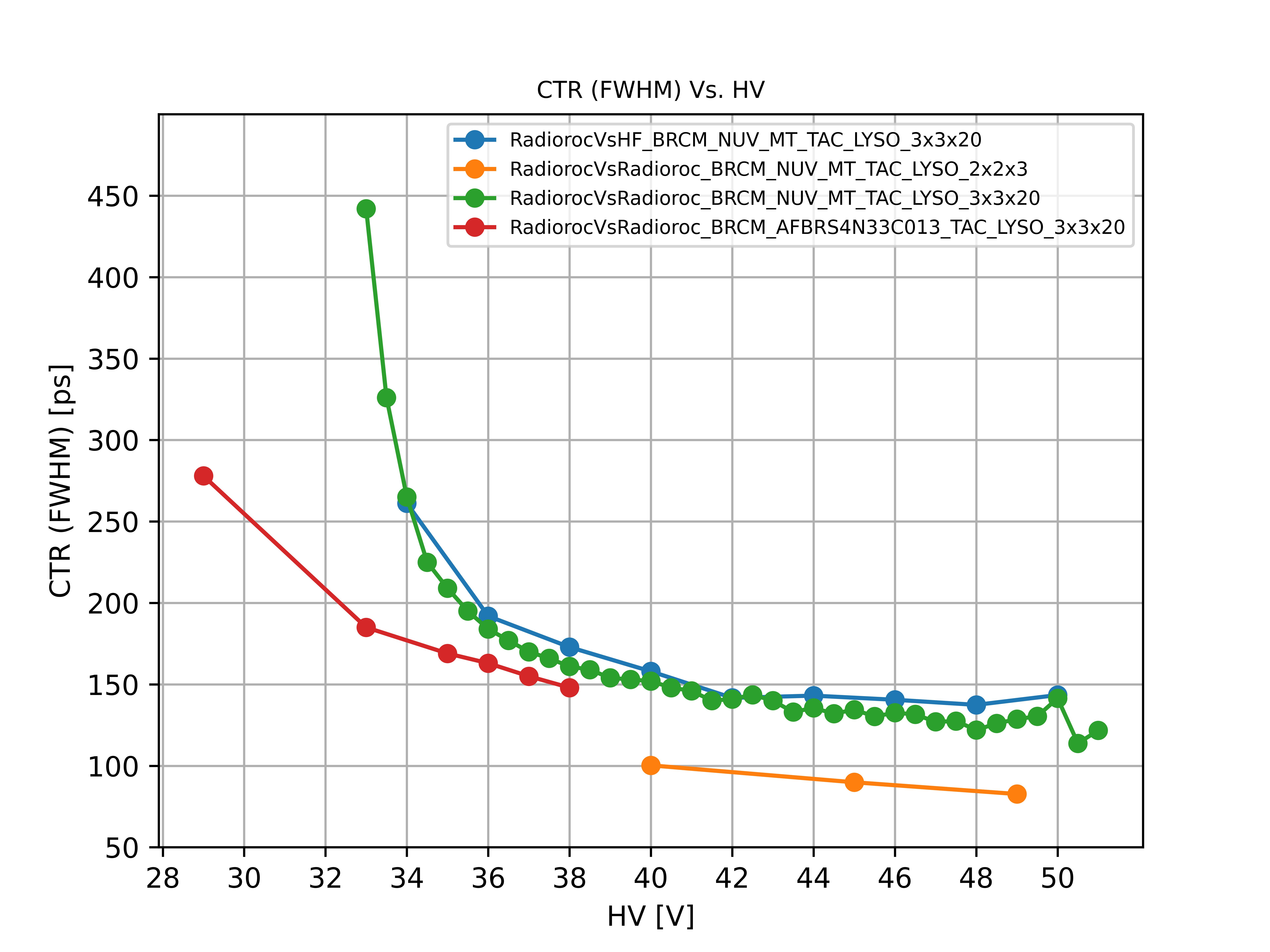}
\caption{Measured CTR of Radioroc with different SiPMs and scintillation crystals.\label{fig:CTR_SummaryPlot}}
\end{figure}

Figure \ref{fig:SPTR_SummaryPlot} and Figure \ref{fig:CTR_SummaryPlot} show, respectively, all single-photon and coincidence time resolution measurements for Liroc and Radioroc with different devices in different configurations. 

The best SPTR values obtained using an Epic black-painted PbF$_{2}$ crystal of 2$\times$2$\times$3 mm$^{3}$ coupled to FBK NUV-HD-LF M3 SiPM: 73 ps (FWHM) for Radioroc and 90 ps (FWHM) for Liroc. The best SPTR measured with the HF readout for the NUV-HD-LF M3 SiPM (shown also in Figure \ref{fig:SPTR_SummaryPlot}) is 27 ps \cite{FBK_MetalMask}. The FBK NUV-HD-RH UHD-DA best SPTR value achieved is 102 ps with Liroc, while the best SPTR for the HF reference readout found with this SiPM is 40 ps.

On the other hand, the best CTR value is 83 ps (FWHM) obtained in a CTR measurement of Radioroc against Radioroc using a small LYSO:Ce,Ca (TAC) of 2$\times$2$\times$\SI{3}{\milli\meter\cubed} coupled to Broadcom NUV-MT SiPM (at a bias voltage of 49 V). Coupling Broadcom NUV-MT SiPMs to larger LYSO:Ce,Ca of 3$\times$3$\times$\SI{20}{\milli\meter\cubed} increases the CTR values from 83 ps (FWHM) (at a bias voltage of 45 V) up to 127 ps (FWHM) at the same configuration.

Scintillator-based detectors are able to achieve very good timing approaching a CTR of 100 ps FWHM, as shown in several publications \cite{6144029, Gundacker_2013}. 
For instance, Broadcom AFBR-S4N33C013 was reported in \cite{SiPM_LeCoq_Gundacker} with intrinsic SPTR lower than 80 ps at an overvoltage of 11 V. The state-of-the-art devices have an FWHM time resolution of around 70 ps \cite{SiPM_LeCoq_Gundacker}. Although the ASIC's time resolution can be limited when coupled with SiPM, the time resolution limits obtained in this study are promising and the current Weeroc's ASICs are achieving at least as best as reference values. 

At the time of writing this article, new versions of these ASICs, with improved functionalities, are being fabricated and will be available soon. In view of this work, Weeroc collaborates with different industrial partners such as CAEN, TAC, and other research institutes to study and optimize a sub-module (ASIC, SiPM, and crystal) for different applications.

\acknowledgments
This project is the result of collaborative work between Weeroc and OMEGA and has received funding from the EC-funded project ATTRACT under grant agreement 777222. The authors would also like to thank Alberto Gola and his team from  Fondazione Bruno Kessler (FBK) for providing the FBK SiPMs to carry out this study. The authors are further grateful to Broadcom Inc. for providing NUV-MT SiPM samples and Mitch Chou from National Sun Yatsen University as well as Jack Lin and Edmund Chin from Taiwan Applied Crystal Co. Ltd. for providing LYSO:Ce,Ca samples.



\bibliographystyle{JHEP} 
\bibliography{biblio}
\end{document}